\documentclass[a4paper,11pt]{article}
\pdfoutput=1
\usepackage{jcappub}

\usepackage{array}
\usepackage{mathtools}

\usepackage{etoolbox}

\title{Topological susceptibility and information theory}
 
\author{C\'esar G\'omez,} 
\emailAdd{cesar.gomez@uam.es}

\affiliation{Instituto de F\'{i}sica Te\'orica UAM-CSIC, Universidad Aut\'onoma de Madrid, Cantoblanco, 28049 Madrid, Spain.}

\date{\today}

\abstract{
In this note we address the question of the $\theta$ dependence in non abelian gauge theories from a pure quantum information point of view. The main result is that the topological susceptibility is the quantum Fisher information of the ground state and that the maximally efficient quantum estimator of $\theta$ can be identified with the physical axion. In this setup the low energy dynamics of the axion is fully determined by quantum estimation theory.}

\begin{document} 
\maketitle
\section{Introduction}

A typical situation in quantum physics is to have observable probability distributions depending on some external free parameters that cannot be directly measured. In this case although the quantum state of the system, for instance the ground state, depends on this set of external parameters we don't have, in the algebra of self adjoint observables, anyone representing them. For a given state, pure or mixed, the amount of information we have about the actual value of these external parameters is encoded in the corresponding quantum Fisher information \footnote{For a good introduction to quantum estimation theory see \cite{Paris} and references therein.}. The simplest and more basic among these external parameters is time itself \footnote{See for instance the discussion in \cite{Aharonov}}. In this case the quantum Fisher information defines a metric on the Hilbert space of states on the basis of which we can define the distinguishability between quantum states at different times \cite{Anandan}. Another example of external parameter, on which we shall focus in this note, is the $\theta$ parameter in QCD \cite{Callan},\cite{Jackiw}. 

In those cases where the external parameter is a gauge artifact i.e. a parameter whose value can be changed using the symmetries of the theory, the corresponding quantum Fisher information is zero. A radical example is time itself in quantum gravity where we can use general covariance to make time a gauge artifact. In this case the corresponding quantum Fisher information vanishes. This situation changes in Cosmology where we include a physical clock defined, for instance, in terms of the inflaton dynamics ( see for instance \cite{R},\cite{B},\cite{S}\cite{Arkani} ).

Perhaps the main lesson we have learnt from our attempts to understand quantum gravity is precisely the absence of external free parameters in the former sense. External parameters should be associated with some dynamical fields and their actual value should be determined dynamically \footnote{The non existence in quantum gravity of free parameters is one of the basic elements of the swampland program \cite{Vafa}. For the case of the axion their necessity, in quantum gravity, was discussed from a microscopic point of view  in \cite{Zell}. The $\theta$ parameter in the string swampland was considered in \cite{VafaC}.} A related issue is the case of superselection charges. In such case we should expect a divergent quantum Fisher information that can be only regularized including non vanishing amplitudes between different superselection sectors i.e. violating the super selection rule itself ( See for references and discussion \cite{Gomez1} ). We will not touch this issue in this note \footnote{A natural question is what is the quantum distance as defined using Bures metric or quantum relative entanglement entropy between different superselection sectors. In case we use to quantify distance between pure states in different superselection sectors the relative entanglement entropy we get infinity.} In the case of the $\theta$ parameter this dynamical field is the well known Peccei Quinn axion field\cite{Peccei}.

What we will suggest here is the simplest possible way to associate with some external parameter a full fledged quantum field. This will be done using the most efficient quantum estimator of the corresponding parameter. This can be done only if we have a finite and non vanishing associated quantum Fisher function. In this way we shall suggest the following conjecture:

{\it In quantum gravity any external parameter, that is not a gauge artifact, is associated with a finite and non vanishing quantum Fisher function.}

In this note we will focus on the $\theta$ parameter of pure Yang Mills theory. The main result is that in this case the quantum Fisher function for the ground state associated with the $\theta$ parameter is fully determined by the topological susceptibility. The associated axion is defined using the maximally efficient quantum estimator of $\theta$ that naturally encodes in its definition the standard theorems defining the low energy dynamics of the axion.


\section{Topological susceptibility}

One of the main non perturbative characterization of the QCD vacuum is the gluon condensate defining the topological susceptibility \cite{Witten}. This condensate parametrizes the $\theta$ dependence of the vacuum energy. For pure $SU(N)$ Yang Mills theory the topological susceptibility is given  by
\begin{equation}
\frac{d^2E(\theta)}{d^2\theta}|_{\theta=0} \equiv C
\end{equation}
with $E(\theta)$ the vacuum energy density.
In terms of the partition function
\begin{equation}
Z(\theta) = \int dA e^{i\int d^4x L(A,\theta)} 
\end{equation}

for $L(A,\theta) = L(A) +\theta \frac{g^2}{16\pi^2 N} F\wedge F$, the topological susceptibility can be formally written as
\begin{equation}
C = \frac{1}{VT} \frac{d^2Z(\theta)}{d^2\theta}|_{\theta=0}
\end{equation}
that up to contact terms reduces to the correlator

\begin{equation}\label{correlator}
C=(\frac{g^2}{16\pi^2 N})^2\int d^4x\langle T( F\wedge F(0) F\wedge F(x)) \rangle
\end{equation}
evaluated at $\theta=0$.

\section{A quantum information description}
Let us denote $\rho_{\theta}$ the density matrix representing the pure state defining the ground state for the corresponding value of the $\theta$ angle. The quantum Fisher information is generically defined as
\begin{equation}
F_q(\theta) = tr(\rho_{\theta} \hat L_{\theta}^2)
\end{equation}
with $\hat L$ defined by
\begin{equation}
\frac{d\rho(\theta)}{d\theta} =\frac{\hat L\rho(\theta) + \rho(\theta) \hat L}{2}
\end{equation}
This quantum Fisher information defines the upper bound of the information we can get to estimate the parameter $\theta$ on the basis of performing arbitrary measurements i.e. it defines the upper bound on the corresponding classical Fisher informations. For $\rho_{\theta}$ a pure state and in the particular case the so defined family of states is a unitary family with generator $G$ the quantum Fisher information is independent on $\theta$ and reduces to
\begin{equation}
F_q = 4 (\Delta G^2)_0
\end{equation}
with $(\Delta G^2)_0 = tr(\rho(0) G^2) - (tr(\rho(0)G))^2$. 

In order to identify the generator $G$ we will use the Hamiltonian description of pure Yang Mills theory in the temporal gauge $A_0=0$ \cite{Witten}. The term in the Hamiltonian linear in $\theta$ is given by
\begin{equation}
H_{1} = \int d^3x \frac{g^2}{16\pi^2N}Tr \pi_iB_i
\end{equation}
for $\pi_i = \partial_0A_i$ and $B_i = \frac{1}{2}\epsilon_{ijk}F_{jk}$. Since the $\theta$ vacua is the ground state for the Hamiltonian $H=H_0+\theta H_1$ we can use standard perturbation theory to define $G$ in terms of $H_1$ leading formally to $H_1= \frac{dG}{dt}$. Using this expression for $G$ we get
\begin{equation}
C= \frac{1}{4VT}F_q
\end{equation}
as the relation between the topological susceptibility $C$ and the quantum Fisher function $F_q$. Note that as done originally in \cite{Witten} we have ignored the contact terms associated with the part of the Hamiltonian depending quadratically on $\theta$. 

Hence we get the basic correspondence:

\vspace{2mm}
{\it Topological susceptibility $\Leftrightarrow$ Quantum Fisher information for $\theta$ parameter}
\vspace{2mm}

We can now use the standard relation between the quantum Fisher function and Bures metric in Hilbert space. The Bures metric satisfies for small $\theta$:
\begin{equation}\label{one}
d_B^2(\rho(\theta)|\rho(0)) = (\Delta G^2)_0 \theta^2 \sim C \theta^2 \sim E(\theta)
\end{equation}

In other words the Bures {\it distance} between two different $\theta$ ground states is determined by the $\theta$ energy $E(\theta)$. 

\subsection{Quantum relative entanglement entropy: small digression}
In addition to Bures metric a very natural candidate to define distances in Hilbert space is the quantum relative entanglement entropy. This distance is defined  for two generic pure states as:
\begin{equation}
S(\rho|\sigma) = tr (\rho ln\frac{\rho}{\sigma})
\end{equation}
By contrast to Bures metric this definition of distance is asymmetric and is not satisfying triangular inequality. In any case both Bures distance as well as relative entanglement entropy are good candidates to define entanglement of a pure state $\rho$ as \cite{Vedral}
\begin{equation}
{\cal{E}}(\rho) = min_{\sigma\in D} d(\rho|\sigma)
\end{equation}
where $D$ is the set of unentangled states and where we can use for $d$ either Bures or relative entanglement entropy notion of distance. In our particular problem this definition allows us to associate with the ground state $\rho(\theta)$ the corresponding measure of entanglement let us say ${\cal{E}}(\theta)$ and to define the corresponding {\it entanglement susceptibility} as $\frac{d^2{\cal{E}}(\theta)}{d^2\theta}$. 

In case the entanglement is defined using Bures metric we can use the triangular inequality and the results of the previous section to prove that the topological susceptibility defines an upper bound on entanglement susceptibility. 

Finally let us note that for both Bures metric as well as quantum relative entanglement entropy we get as the first law of entanglement the positivity of the metric. This in particular means that $d^2_B(\rho(\theta)|\rho(0))$ has a minimum for $\theta=0$ which is in this context the analog of the well known Vafa-Witten theorem \cite{VafaWitten}.

\section{Axion as a quantum estimator}
Let us think on $\theta$ as a parameter on which the different physical correlators depend. The quantum Fisher information or equivalently as shown above the topological susceptibility of the vacuum encodes how much information we have to estimate the physical value of $\theta$. As it is standard in estimation theory we can define a quantum estimator for this parameter. Around $\theta=0$ the quantum estimator is simply
\begin{equation}\label{def}
\hat A(x) =  \frac{G(x)} {F_q}
\end{equation}
where by $G(x)$ we mean the density of the generator $G$. As it is defined this operator has zero dimension. Let us define the  associated axion as $\phi \equiv f_a \hat A$ for $f_a$ with units of mass.

Before going into more details let us recall some basic properties of quantum estimators. Generically, in quantum estimation theory, the parameter under study is not {\it the eigenvalue of any self adjoint operator} and therefore we don't have in the algebra of observables anyone corresponding with the direct measurement of the parameter. By contrast a {\it quantum estimator} is a self adjoint operator that describes a quantum measurement {\it followed by a classical processing of the data}. The most efficient classical data processing is determined by the maximal amount of information given by the quantum Fisher function. Heuristically the logic underlying this definition of quantum estimator is to identify a self adjoint operator with the same quantum uncertainty you expect for the would be operator conjugated to the generator $G$. This follows from the relation $\Delta (\hat A^2) = \frac{1}{F_q}$.
Thus the corresponding two point function for the so defined axion field, at zero momentum, is given by
\begin{equation}\label{propagator}
\Delta(\phi^2) = \frac{f_a^2}{F^2}\langle \hat L_k \hat L_0 \rangle |_{k=0} = \frac{f_a^2}{F_q}
\end{equation}
giving an effective mass for $\phi$
\begin{equation}
m^2 = \frac{F_q}{f_a^2} = \frac{C}{f_a^2}
\end{equation}
The very definition of the axion field $\phi$ as the quantum estimator of $\theta$ encodes the well known {\it low energy theorems}. In fact from (\ref{def}) it follows
\begin{equation}
\langle \phi |G|0\rangle= \frac{F_q}{f_a}
\end{equation}
Using the form of $G$ and $F_q$ this is the standard low energy theorem defining the effective coupling between the axion field $\phi$ and $F\wedge F$. In this form the topological susceptibility can be written, using (\ref{propagator}), as
\begin{equation}
\langle F\wedge F F\wedge F\rangle = \langle F\wedge F |\phi\rangle \Delta^2(\phi) \langle \phi|F\wedge F\rangle =
\frac{F_q^2}{f_a^2}. \frac{f_a^2}{F_q} =  F_q = C
\end{equation}
In summary we get the following two basic correspondences:

\vspace{2mm}
{\it Axion $\Leftrightarrow$ Quantum estimator of $\theta$ parameter}
\vspace{2mm}

and

\vspace{2mm}
{\it Axion low energy theorems $\Leftrightarrow$ Axion as a maximally efficient quantum estimator}
\vspace{2mm}

It is important to stress the differences between the former approach and the one initially developed by Witten in the solution to the $U(1)$ problem. In that case the fermionic contribution, in the case of massless fermions, to the Fourier transform of the topological susceptibility $U(k) = \int d^4x e^{ikx}T(F\wedge F(x) F\wedge F(0))\rangle$ should compensate the pure Yang Mills contribution. Denoting $C$ the pure Yang Mills topological susceptibility this implies the existence of the $\eta'$ meson contributing to $U(k=0)$ as
$\frac{c_{\eta'}^2}{m^2_{\eta'}}$ leading to the basic formula \cite{Witten}\cite{Veneziano}
\begin{equation}
m^2_{\eta'} = \frac{c_{\eta'}^2}{C}
\end{equation}
The low energy theorem is now represented as
\begin{equation}
c_{\eta'}^2 = m_{\eta'}^4 f_a^2
\end{equation}
leading to 
\begin{equation}
m_{\eta'}^2 = \frac{C}{f_a^2}
\end{equation}
In our approach and for pure Yang Mills we define the quantum estimator operator with units of mass that we introduce by hand using $f_a$. Once this quantum estimator $\phi$ is defined everything is determined, namely $\langle \phi|\hat L\rangle$ as well as $\langle \phi_k\phi_0\rangle$ in the $k=0$ limit. In other words the constraint used in \cite{Witten} to relate $U(k=0)$ and $c_{\eta'}^2$ using the low energy theorems is automatically implemented for the quantum estimator that plays the role of the effective axion field. What we normally discover in massless QCD is that the $\eta'$ meson plays the role of the quantum estimator of $\theta$. This is very natural since the operator defining the quantum estimator is the one defining transformations of $\theta$ that in the case of massless fermions is just the $U(1)$ axial charge.
Of course in this case the scale $f_a$ is just the analog of $f_{\pi}$ for the corresponding Goldstone boson. In the case we have not available massless fermions to define $\hat L$ the corresponding quantum estimator is an extra ingredient which is nothing else but the Peccei Quinn axion for the solution of the strong CP problem. In this case $f_a$ is completely undetermined by the low energy dynamics. 

Summarizing, the {\it quantum estimation relation} for the $\theta$ parameter
\begin{equation}
\hat \phi = \frac{G f_a}{F_q}
\end{equation}
becomes the low energy theorem
\begin{equation}
\partial_{\mu}J^{\mu}_{PQ} = m^2 f_a \hat \phi
\end{equation}
for the PQ current with $m^2= \frac{F_q}{f_a^2}$ \footnote{Before ending we shall like to make a general cosmological comment. In the case of natural inflation the former quantum information approach to the axion can be combined with the recent quantum information approach to cosmology \cite{qf1}. For such an axion candidate to natural inflation \cite{Inflation} we could expect the relation
\begin{equation}
\frac{f_a^2}{F_q} \sim \frac{1}{M_P^2 \epsilon_a^2}
\end{equation}
for $\epsilon_a$ the slow roll parameter for natural ( axionic ) inflation. Using this relation between the quantum Fisher information of the $\theta$ angle associated with the axionic inflaton and the cosmological Fisher information defined in terms of slow roll parameters we get as the estimation of $f_a$
\begin{equation}
f_a = (\frac{\Lambda}{M_P})^2 \frac{1}{\epsilon_a} M_P
\end{equation}
where we have used $F_q \sim \Lambda^4$. Thus for $\Lambda$ defined at the GUT grand unified scale $10^{16} Gev$ and for a value of $\epsilon_a$ based on imposing a reasonable amount of e-foldings $O(10^3)$ we get 
$f_a \sim \Lambda_{GUT} = 10^{16} Gev$.}.

The logic underlying the philosophy presented in this note can be easily encapsulated. Processing information by means of the quantum Fisher function provides the natural procedure to define a self adjoint operator that, for those practical purposes concerning quantum uncertainty, is equivalent to the a priori non existent, in the algebra of observables, operator measuring the $\theta$ angle. Whenever we we have a local density for the generator $G$ the quantum estimator becomes a local field whose action on the Hilbert space is non linearly defined. The non linearity lying in the classical information processing. The low energy dynamics of the so defined local field is then fully determined by its definition as quantum estimator. In particular, as stressed above, in the case we define a formal PQ current, the corresponding low energy theorem is set by the quantum Fisher information:
\begin{equation}
f_a \langle 0|\partial_{\mu}J^{\mu}_{PQ}|\phi\rangle = F_q
\end{equation}
In general we can have different realizations of $J_{PQ}$ depending on how we complete the theory. In the case of massless QCD , the current is just the $U(1)$ axial current and is the mass of the $\eta'$ what encodes the quantum Fisher information about the $\theta$ angle.

Summing up axion dynamics can be thought as arranging, by means of quantum estimators, maximal quantum information about the corresponding $\theta$ angles.

\acknowledgments
 This work was supported by grants SEV-2016-0597, FPA2015-65480-P and PGC2018-095976-B-C21.


\begin{thebibliography}{99}
  \bibitem{Paris}
 M.~G.~A.~ Paris,
 International Journal of Quantum Information, 2009 -
  \bibitem{Aharonov}
Y.~Aharonov and D.~Bohm,
Phys. Rev. \textbf{122} (1961) no.5, 1649-1658
\bibitem{Anandan}
J.~Anandan and Y.~Aharonov,
Phys. Rev. Lett. \textbf{65} (1990), 1697-1700
\bibitem{Callan}
C.~G.~Callan, Jr., R.~F.~Dashen and D.~J.~Gross,
Phys. Lett. B \textbf{63} (1976), 334-340
\bibitem{Jackiw}
R.~Jackiw and C.~Rebbi,
Phys. Rev. Lett. \textbf{37} (1976), 172-175
   
\bibitem{R}V. G. Lapchinsky and V. A. Rubakov, 
 Acta Phys. Polon. B10, 1041 (1979).

\bibitem{B}T. Banks, 
 Nucl.Phys. B249, 332 (1985).

\bibitem{S} T. Banks, W. Fischler and L. Susskind, 
 Nucl. Phys. B262, 159 (1985).
\bibitem{Arkani}
N.~Arkani-Hamed, S.~Dubovsky, A.~Nicolis, E.~Trincherini and G.~Villadoro,
JHEP \textbf{05} (2007), 055

[arXiv:0704.1814 [hep-th]].
\bibitem{Vafa}T. D. Brennan, F. Carta and C. Vafa, 
 arXiv:1711.00864 [hep-th].
 
\bibitem{Zell}
G.~Dvali, C.~Gomez and S.~Zell,
[arXiv:1811.03079 [hep-th]].
\bibitem{VafaC}
S.~Cecotti and C.~Vafa,
[arXiv:1808.03483 [hep-th]].
\bibitem{Peccei}
R.~D.~Peccei and H.~R.~Quinn,
Phys. Rev. Lett. \textbf{38} (1977), 1440-1443
\bibitem{Gomez1}
C.~Gomez,
[arXiv:1907.03619 [hep-th]].
\bibitem{Inflation}
K.~Freese, J.~A.~Frieman and A.~V.~Olinto,
Phys. Rev. Lett. \textbf{65} (1990), 3233-3236
\bibitem{Witten}
E.~Witten,
Nucl. Phys. B \textbf{156} (1979), 269-283
\bibitem{relative1} 
M. Ohya and D.   Petz,  "Quantum entropy and its use",  Text and Monographs in Physics, Springer Study Edition, Corrected 2nd Printing, 2004. \bibitem{relative2} 
H. Araki,  Publ. Res. Inst.Math. Sci. Kyoto1976, 809 (1976)

\bibitem{Vedral}
V.~Vedral and M.~B.~Plenio,
Phys. Rev. A \textbf{57} (1998), 1619-1633
doi:10.1103/PhysRevA.57.1619
[arXiv:quant-ph/9707035 [quant-ph]].
\bibitem{VafaWitten}
C.~Vafa and E.~Witten,
Nucl. Phys. B \textbf{234} (1984), 173-188

\bibitem{Veneziano}
G.~Veneziano,
Nucl. Phys. B \textbf{159} (1979), 213-224

\bibitem{qf1}
C.~Gomez and R.~Jimenez,
[arXiv:2002.04294 [hep-th]].


C.~Gomez and R.~Jimenez,
[arXiv:2003.08402 [astro-ph.CO]].

C.~Gomez and R.~Jimenez,
[arXiv:2005.09506 [astro-ph.CO]].
%
\end{thebibliography}
\end{document}